\documentclass[sigconf]{acmart}
\usepackage{booktabs} % For formal tables

\usepackage{booktabs} % For formal tables
\usepackage{amssymb}
\usepackage{amsmath}
\usepackage{bm}
\usepackage{subcaption}
\usepackage{url}
 
\usepackage{cancel}
\usepackage[ruled,vlined,linesnumbered]{algorithm2e}

\SetKwInput{KwInput}{Input}
\SetKwInput{KwOutput}{Output}

\setcopyright{acmcopyright}
\copyrightyear{2018}
\acmYear{2018}
\setcopyright{acmcopyright}
\acmConference[CIKM '18]{The 27th ACM International Conference on Information and Knowledge Management}{October 22--26, 2018}{Torino, Italy}
\acmBooktitle{The 27th ACM International Conference on Information and Knowledge Management (CIKM '18), October 22--26, 2018, Torino, Italy}
\acmPrice{15.00}
\acmDOI{10.1145/3269206.3271746}
\acmISBN{978-1-4503-6014-2/18/10}
\begin{document}
\title{ Signed Network Modeling Based on Structural Balance Theory}
% \titlenote{Produces the permission block, and
% \subtitle{Extended Abstract}
% \subtitlenote{The full version of the author's guide is available as
%   \texttt{acmart.pdf} document}

% \author{Anonymous Author(s)}
% \affiliation{%
%   \institution{Affiliation(s)\\Affiliation(s)}
% }
%  \email{Email(s)}

\author{Tyler Derr}
\affiliation{%
  \institution{Data Science and Engineering Lab\\Michigan State University}
}
 \email{derrtyle@msu.edu}

\author{Charu Aggarwal}
\affiliation{%
  \institution{IBM T.J. Watson Research Center}
}
 \email{charu@us.ibm.com}

\author{Jiliang Tang}
\affiliation{%
  \institution{Data Science and Engineering Lab\\Michigan State University}
}
 \email{tangjili@msu.edu}

\begin{abstract}
The modeling of networks, specifically generative models, has been shown to provide a plethora of information about the underlying network structures, as well as many other benefits behind their construction. There has been a considerable increase in interest for the better understanding and modeling of networks, and the vast majority of existing work has been for unsigned networks. However, many networks can have positive and negative links (or signed networks), especially in online social media. It is evident from recent work that signed networks present unique properties and principles from unsigned networks due to the added complexity, which pose tremendous challenges on existing unsigned network models.  Hence, in this paper, we investigate the problem of modeling signed networks. In particular, we provide a principled approach to capture important properties and principles of signed networks and propose a novel signed network model guided by Structural Balance Theory.  Empirical experiments on three real-world signed networks demonstrate the effectiveness of the proposed model. 
\end{abstract}




\maketitle

\section{Introduction}
Network modeling aims to design a model to represent a complex network through a few relatively simple set of equations and/or procedures such that, when provided a network as input, the model can learn a set of parameters to construct another network that is as similar to the input as possible. Ideally this would result in many observable/measurable properties being maintained from the input to the generated output network. In unsigned networks, the typical modeled properties are the power law degree distribution~\cite{chung2002average,leskovec2005graphs, leskovec2010kronecker, barabasi1999emergence}, assortativity~\cite{mussmann2015assortativity,newman2002assortative}, clustering coefficients~\cite{pfeiffer2012fast,watts1998collective,seshadhri2012community,holme2002growing}, and small diameter~\cite{leskovec2005graphs,watts1998collective}. Nowadays, more data can be represented as large networks in many real-world applications such as the Web~\cite{szell2010multirelational,leskovec2009community}, biology~\cite{barabasi2004network,volz2007susceptible}, and social media~\cite{malekzadeh2011social,leskovec2012learning}. Increasing attention has been attracted in better understanding and modeling networks. Traditionally network modeling has focused on unsigned networks. However, many networks can have positive and negative links (or signed networks~\cite{cartwright1956structural,heider1946attitudes,tang2015survey}), especially in online social media, which then raises the question -- whether dedicated efforts are needed to model signed networks in addition to the unsigned techniques.

Signed networks are unique from unsigned not only due to the increased complexity added to the network by having a sign associated with every edge, but also (and more importantly) because there are specific principles (or social theories), such as balance theory, that play a key role driving the dynamics and construction of signed networks~\cite{leskovec2010signed,leskovec2010predicting,derr2018opinions}. For example, in unsigned networks we have the property of transitivity and we see a large amount of local clustering (i.e., formation of triangles). In comparison, with signed networks, not only are their patterns in the network driving local clustering, but also in the distribution of triangles (based on their edge signs) found in the network. Suggested by balance theory \cite{heider1946attitudes}, some triangles are more likely to be formed (i.e., balanced) than others (i.e., unbalanced) in signed networks. Hence, modeling signed networks requires to preserve not only unique properties of signed networks such as the sign distribution, but also other properties suggested by their principles such as the distribution of formed triangles. However, these mechanisms are not incorporated into unsigned network modeling and unsigned network models are unequipped for signed networks. Thus, there is a need to design network models for signed networks. 

Network models have many direct applications and a diverse set of benefits beyond and including the better understanding of the network structure and dynamics. Currently there is a significant push for better anonymization in social media. However, for researchers wanting to further advance their field, it is necessary to utilize the network data for knowledge discovery, mining, and furthermore for testing and benchmarking their methods and algorithms. A generative network model could be utilized for constructing synthetic networks having similar properties as their corresponding real network, but without compromising the user's privacy and allowing further advancements through the use of the synthetic network datasets. Similarly such a model can be used as a null-model for network property significance testing or for constructing synthetic networks of varying network properties to further understand the relationship between the network model and real world networks in terms of their dynamics and construction process.

In this paper, we aim to investigate the problem of signed network modeling.  We propose a novel signed network model, which targets to preserve three key properties of signed networks -- (1) degree distribution; (2) sign distribution; and (3) balance/unbalanced triangle distribution suggested by balance theory. Our contributions are summarized as follows:
\begin{itemize}
\item Introduce a principled approach to capture key properties of signed networks into a coherent model;
\item Propose a novel generative signed network model, which provides a generative process and automated parameter learning algorithms to preserve key properties of signed networks simultaneously; and 
\item Conduct experiments on three real-world signed networks to demonstrate the effectiveness of the proposed framework in generating signed networks.
\end{itemize}

The rest of this paper is organized as follows. In Section 2, we formally introduce the generative signed network modeling problem. Then we detail the proposed model with a generative process and parameter learning algorithms in Section 3. Section 4 presents the experimental results. In Section 5, we review related work. Conclusions and future work are given in Section 6.

\section{Problem Statement}

A signed network $\mathcal{G}$ is composed of a set $V = \{v_1, v_2, \dots, v_N\}$ of $N$ vertices, a set of $M^+$ positive links $\mathcal{E}^+$ and a set of $M^-$ negative links $\mathcal{E}^-$. Let $\mathcal{E} = \mathcal{E}^+ \cup \mathcal{E}^-$ represent the set of $M = M^+ + M^-$ edges in the signed network when not considering the sign. Note that in this work, we focus on undirected and unweighted signed networks and would like to leave modeling directed and weighted signed networks as future work.

We can formally define the generative signed network modeling problem as follows: \\
\textit{Given a signed network $\mathcal{G}_I=(V_I,\mathcal{E}^+_I,\mathcal{E}^-_I)$ as input, we seek to learn a set of parameters $\Theta$ for a given model $\mathcal{M}$ that can retain the network properties found in $\mathcal{G}_I$, such that we can construct synthetic output networks $\mathcal{G}_o=(V_o,\mathcal{E}^+_o,\mathcal{E}^-_o)$, using $\mathcal{M}$ based on $\Theta$, that closely resemble the input network in terms of measured network properties. }

\section{The Proposed Signed Network Model}
Traditionally network modeling has focused on unsigned networks and preserving unsigned network properties. Signed networks have distinct properties from unsigned networks~\cite{tang2014distrust,beigi2016signed}. For example, negative links are available in signed networks and ignoring the negative links can result in over-estimation of the impact of positive links \cite{li2013influence}; and most of the triangles in signed networks satisfy balance theory \cite{szell2010multirelational}.  However, these properties cannot be simply captured by unsigned network models. Hence, dedicated efforts are demanded to model signed networks. In this section, we will introduce the proposed signed network model. The notations we will use in the remainder of the paper are demonstrated in Table~\ref{tab:notations}. 

\begin{table}
\begin{center}
\caption{Notations.}
\label{tab:notations}
\resizebox{\columnwidth}{!}{
\begin{tabular}{|l|l|}
	\hline
	\label{tab:notations} 	 
	Notations & Descriptions\\
	\hline
	$e_{ij}$ & undirected edge between vertices $v_i$ and $v_j$\\
	$N_{i}$ & set of neighbors for node $i$ \\
	${\bf d}$ & degree vector based on $\mathcal{E}_I$ where $d_i$ is the degree of $v_i$\\
	$\eta$ & fraction of links being positive in $\mathcal{G}_I$ (i.e., $M_I^+ / M_I$)\\
	$\Delta_B$ & fraction of triangles balanced in $\mathcal{G}_I$\\
	$\bm{\pi}$ & sampling vector from degree distribution in $\mathcal{E}_I$\\
	$\rho$ & probability new edge closes a wedge to be a triangle in $\mathcal{G}_o$ \\
	& via two-hop walk instead of a randomly inserting an edge\\
	$\alpha$ & probability a randomly inserted edge into $\mathcal{G}_o$ is positive\\	
	$\beta$ & probability of closing a wedge to have more balanced triangles in $\mathcal{G}_o$\\
	$\Delta_{ij(B)}^{random}$ & approximation to the $\mathbb{E}$[\# of (balanced) triangles]\\
	& that will get created due to randomly inserting $e_{ij}$ \\
	$\Delta_{\mathcal{G}(B)}^{random}$ & average of $\Delta_{ij(B)}^{random}$ across all possible edges\\
	$\Delta_{ij(B)}^{triangle}$ & approximation to the $\mathbb{E}$[\# (balanced) triangles] \\
	& created when inserting edge $e_{ij}$ via the wedge closing procedure\\	
	$\Delta_{\mathcal{G}(B)}^{triangle}$ & average of $\Delta_{ij(B)}^{triangle}$ across all possible edges\\
	\hline
\end{tabular}}
\vspace{-0.2in}
\end{center}
\end{table}

\subsection{Balanced Signed Chung-Lu Model}
Previous studies have demonstrated that the node degrees of signed networks also follow power law distributions \cite{tang2014distrust,derr2018relevance} similar to that of unsigned networks.  Hence, we propose to build the signed network model based on the unsigned Chung-Lu model, which can preserve the degree distributions of the input work. The Chung-Lu (CL) model first takes an unsigned network $\mathcal{G}_I=(V_I,\mathcal{E}_I)$ as input and independently decides whether each of the $N^2$ edges are placed in the generated network with each edge $e_{ij}$ having probability $\frac{d_i d_j}{2M}$ where $d_i$ is the degree of node $v_i$ and $M$ is the number of edges in the network. It can be shown that the expected degree distribution of the output network $\mathcal{G}_o$ is equivalent to that of $\mathcal{G}_I$. A fast variant of the Chung-Lu model, FCL \cite{pinar2012similarity}, is proposed to create a vector $\bm{\pi}$ which consists of $2M$ values, where for each edge both incident vertices are added to the vector. Rather than deciding whether each of the $N^2$ edges get added to the network (as done in CL), FCL can just randomly sample two vertices from $\bm{\pi}$ uniformly, since this simulates the degree distribution. Note that FCL ignores self-loops and multi-edges when sampling $M$ edges. However, since most real-world unsigned networks have higher clustering coefficients than those generated by CL and FCL, another CL variant Transitive Chung-Lu (TCL) was introduced in \cite{pfeiffer2012fast} to maintain the transitivity. Rather than always picking two vertices from $\bm{\pi}$, instead, TCL occasionally picks a single vertex from $\bm{\pi}$ and then, with a parameter $\rho$,  performs a two-hop walk to select the second vertex. When including this edge, the process is explicitly constructing at least one triangle by closing off the wedge (i.e., wedge closing procedure) created by the two-hop walk. 

The proposed Balanced Signed Chung-Lu model (BSCL) is based on the TCL model, which automatically allows the mechanism for closely approximating the degree distribution and also the local clustering coefficient during the construction process. However, as previously mentioned, the distribution of formed triangles is a key property in signed networks and most of these triangles adhere to balance theory. Note that, when performing the wedge closure procedure, we are not only closing the single wedge we explicitly constructed (through our two-hop walk), but there could be other common neighbors between these two vertices. Thus, we introduce a parameter $\beta$, which denotes the probability of assigning the edge sign to ensure the majority of the triangles being created by this new edge are balanced. With the introduction of this parameter, our model is able to capture a range of balance in signed networks. This is necessary since not all signed networks are completely balanced, and in fact real-world networks can have a varied percentage of triangles being balanced~\cite{leskovec2010predicting}. 

Meanwhile, we also want to maintain the sign distribution. However, the above process of determining the edge sign for wedge closure is based on balance theory (i.e., local sign perspective) and not on the global sign perspective (i.e., $\eta$). This implies that when randomly inserting an edge into the network if we simply choose the sign based on $\eta$, then this could lead to our generated networks deviating from the true sign distribution of the input network. Therefore, we introduce $\alpha$, which is a corrected probability (instead of using $\eta$) for a randomly inserted link and is used to correct the bias of positive or negative edges that will be created through the use of $\beta$ (which is from the local sign perspective).

With the introduction of three parameters (i.e., $\rho$, $\alpha$ and $\beta$), the proposed balanced signed Chung-Lu model (BSCL) is shown in Algorithm \ref{alg:BSCL}. Here we step through the high level processes of BSCL before later discussing both the network generation process and the parameter learning algorithms. On line 1 of Algorithm~\ref{alg:BSCL}, we first construct $\mathcal{E}$. Then, using the degree distribution of $\mathcal{E}$, we can construct the vertex sampling vector $\bm{\pi}$ as shown on line 2. Next we calculate the properties of the input network we aim to preserve. These include the percentage of positive links $\eta$, vector of degrees $\mathbf{d},$ and percentage of balanced triads $\Delta_B$ from lines 3 to 5.  With these values, we will estimate the major parameters of BSCL including $\rho$, $\alpha$ and $\beta$ as mentioned on line 6 using our learning algorithms that will be discussed in subsection \ref{subsec:param_learning}. Finally, we generate the network based on the learnt parameters on line 7 and then output the constructed synthetic signed network $\mathcal{G}_o$. In the next subsection, we will discuss the details of the network generation process performed by BSCL and then discuss how these parameters can be automatically and efficiently learned.

\subsection{Network Generation for BSCL}

Given the parameter values for $\rho, \alpha$ and $\beta$, we show in Algorithm~\ref{alg:BSCL_Network_Generation} how BSCL can generate a synthetic signed network maintaining the key signed network properties. First, on line 1, we use the FCL method for the construction of a set $\mathcal{E}_o$ of M edges, which closely approximates to the original degree distribution. Then, on line 2, we split the unsigned edges into two sets, $\mathcal{E}^+_o$ and $\mathcal{E}^-_o$, by randomly assigning edge signs, based on $\eta$, such that the percentage of positive links matches that of the input network. Next, from lines 3 to 17, we add $M$ new edges to the network, one at a time, while removing the oldest edge in the network for each new edge inserted. The reason for starting with this initial set of edges from FCL on line 1 is due to the fact when performing the wedge closing procedure (from lines 5 to 10), if the starting network is initially too sparse, there will not be many opportunities for two-hop walks to create triangles. We note that after each iteration from lines 3 to 17, $\mathcal{G}_o$ maintains the correct total number of edges, $M$.

We will either insert an edge by closing a wedge into a triangle and using $\beta$ to help maintain the balance, or insert a random edge and select its sign based on $\alpha$ to correctly maintain the sign distribution. On line 5, we use our parameter $\rho$ to determine which edge insertion method we will use. Next we will further discuss these two edge insertion procedures.

\begin{algorithm}[]%H	
\DontPrintSemicolon
\KwInput{Signed Network $\mathcal{G}_I=(V_I,\mathcal{E}^+_I,\mathcal{E}^-_I)$}
\KwOutput{Synthetic Signed Network $\mathcal{G}_o=(V_I,\mathcal{E}^+_o,\mathcal{E}^-_o)$}
$\mathcal{E}_I = \mathcal{E}^+_I \cup \mathcal{E}^-_I$\;

$\bm{\pi} \leftarrow$ Sampling vector based on the degree distribution in $\mathcal{E}_I$\;

$\eta \leftarrow \frac{M^+_I}{M^+_I + M^-_I}$  \;

$\mathbf{d} \leftarrow $ Calculate\_Degree\_Vector($V_I,\mathcal{E}_I$)

$\Delta_B \leftarrow $ Percentage balanced triangles in $\mathcal{G}_I$\;

$\rho, \alpha, \beta \leftarrow$ Parameter\_Learning($\mathcal{E}_I, \eta,\Delta_B,\mathbf{d},M$\textbf{)}\;

$\mathcal{E}^+_o,\mathcal{E}^-_o \leftarrow$ Network\_Generation($\eta,M,\bm{\pi},\rho,\alpha,\beta$\textbf{)}\;
\caption{Balanced Signed Chung-Lu Model\label{alg:BSCL}}
\end{algorithm}

\begin{algorithm}[]%H	
\DontPrintSemicolon
		$\mathcal{E}_o = FCL($M,$\bm{\pi}$)\;
		
		$\mathcal{E}^+_o,\mathcal{E}^-_o \leftarrow $ Randomly partition $\mathcal{E}_o$ based on $\eta$
		
		\For{1 to M}{
			$v_i$ = sample from $\bm{\pi}$\;
			
			\If{wedge\_closing\_procedure$(\rho)$}{
				$v_j$ = Perform two-hop walk from $v_i$ through neighbor $v_k$\;
				
				\If{close\_for\_balance($\beta$)}{
					Add $e_{ij}$ to $\mathcal{E}^+_o$ or $\mathcal{E}^-_o$ based on the sign that closes the wedge and other common neighbors to have more balanced triangles
				}
				\Else{
					Add $e_{ij}$ to $\mathcal{E}^+_o$ or $\mathcal{E}^-_o$ to have more unbalanced triangles
				}
			}
			\Else(insert a random edge){
				$v_j$ = sample from $\bm{\pi}$\;
				
				\uIf{create\_positive\_edge$(\alpha)$}{
					$\mathcal{E}^+_o \leftarrow \mathcal{E}^+_o \cup \{e_{ij}\} $\;
				}
				\Else{
					$\mathcal{E}^-_o \leftarrow \mathcal{E}^-_o \cup \{e_{ij}\} $\;
				}
			}
			Remove oldest edge from $\{\mathcal{E}^+_o \cup \mathcal{E}^-_o\}$ respectively		
		}
		
		\Return{		$\mathcal{E}^+_o,\mathcal{E}^-_o$\;
}		
		\caption{Network\_Generation($\eta, M,\pi, \rho, \alpha, \beta$) \label{alg:BSCL_Network_Generation}}
\end{algorithm}

The wedge closing procedure is selected with probability $\rho$ on line 5, but starts on line 4 with the selection of $v_i$ uniformly at random from $\bm{\pi}$. Then, on line 6, we perform a two-hop walk from $v_i$ through a neighbor $v_k$ to land on $v_j$. We have just selected the wedge consisting of edges $e_{ik}$ and $e_{kj}$ to close into a triangle. We note that although we are explicitly constructing the triangle composed of vertices $v_i$, $v_k$ and $v_j$ that edge $e_{ij}$ would also implicitly be closing wedges to form triangles with any other common neighbors that $v_i$ and $v_j$ might have. Hence, we use our learned parameter $\beta$ for determining if we should introduce more balanced or unbalanced triangles into the network based on the total balance in the input signed network (i.e., $\Delta_B$). Therefore, on line 7, with probability $\beta$, we choose to select the edge sign of $e_{ij}$ such that the majority of the triangles being created (both those explicitly through the two-hop walk and implicitly through other common neighbors) will adhere to balance theory. As mentioned on line 8, depending on whether balance theory would suggest $e_{ij}$ to being positive or negative, we will add the edge to the set $\mathcal{E}_o^+$ or $\mathcal{E}_o^-$, respectively. Similarly, with probability (1-$\beta$), the sign of $e_{ij}$ will be selected to introduce more unbalanced triangles into the generated network.

If not performing the wedge closing procedure, BSCL will instead insert a random edge with probability $(1-\rho)$. This process starts similarly as line 4 by selecting the first vertex $v_i$. Then, on line 12, a second vertex is sampled from $\bm{\pi}$ such that we can then insert edge $e_{ij}$ into the network. However, since we desire our generated network to maintain the correct sign distribution, the sign for the edge $e_{ij}$ needs to carefully be determined. As previously discussed, the wedge closing procedure will disrupt the global sign distribution and therefore rather than using $\eta$ for the sign selection, we use our learned parameter $\alpha$. We again note that $\alpha$ will be learnt such that it incorporates the bias induced from the local sign selections made during the wedge closing procedure controlled by $\beta$. Therefore, with probability $\alpha$, on line 13, we choose to go to line 14 and insert $e_{ij}$ as a positive link and add it to the set $\mathcal{E}_o^+$. On the other hand, with probability $(1-\alpha)$, we go to line 16 and select $e_{ij}$ to be negative and therefore add it to the set of negative edges $\mathcal{E}_o^-$.

After edge insertion, the next step is to remove the oldest edge in the generated network $\mathcal{G}_o$ such that it maintains $M$ edges. Line 17 shows that we select the oldest edge from the union of the positive and negative edge sets (i.e., ${\mathcal{E}_o^+ \cup \mathcal{E}_o^-}$) and then respectively remove it from the edge set it was selected from. After performing this loop from lines 3 to 17 $M$ times, all the initial edges from FCL will have been removed and the network generator can return the resulting positive and negative edge sets $\mathcal{E}_o^+$ and $\mathcal{E}_o^-$, respectively.

One step we did not mention in Algorithm \ref{alg:BSCL_Network_Generation} for ease of description is that we also make use of a queue for when having collisions (i.e, selecting to insert an edge that already exists in the network or a self-loop). For every time we have such a collision, the vertices are added to the queue. Then, before each time selecting an edge from $\bm{\pi}$ (i.e. on lines 4 and 12), the queue is checked. If the queue is empty, then we proceed to sample from $\bm{\pi}$. However, if the queue is non-empty, then we instead take from the front of the queue. Similarly, we utilize the queue if unable to perform a two hop walk from vertex $v_i$. Next we will discuss how we can learn the parameters $\rho, \alpha, $ and $\beta$.

\vspace{-2ex}
\subsection{Parameter Learning} \label{subsec:param_learning} 
In the last subsections, we have introduced the BSCL model and network generation process based on the parameters $\rho, \alpha, $ and $\beta$, here we discuss how to learn these parameters from the input signed network. We notice that these parameters are related to each other. For example, when constructing triangles to be balanced or unbalanced (based on $\beta$), this will disrupt the global sign distribution since these decisions are only based on the local sign perspective. Similarly, when inserting a random edge with a sign based on $\alpha$, this has the potential to disrupt the distribution of triangles and the percentage of triangles that are balanced in the network. This is because the decision for the sign of a random edge insertion is based solely on the global sign perspective and ignores the local perspective of whether triangles are being created via this inserted edge to be balanced or unbalanced. Hence, next we discuss the proposed algorithm for learning these parameters alternatively and iteratively.

\vspace{-2ex}
\subsubsection{Learning $\rho$.}
For the parameter $\rho$, we make use of the Expectation-Maximization (EM) learning method following a similar process in the TCL model~\cite{pfeiffer2012fast}. The general idea is that it can be learned after defining a hidden variable associated with each edge, which determine whether the edge was added to the network randomly or through a wedge being closed into a triangle. More specifically,  let $z_{ij} \in Z$ be the latent variable assigned to each edge $e_{ij}$. These latent variables can be equal to 1 or 0, where $z_{ij} = 0$ indicates that the edge was created via random sampling from $\bm{\pi}$ and $z_{ij} =1$ suggests that the edge $e_{ij}$ was created through the two-hop walk wedge closing procedure.

Let $\bm{\pi}_i$ represent the probability of selecting $v_i$ from the sampling vector $\bm{\pi}_i$, $\mathbb{I}[v_j \in N_k]$ as an indicator function to be 1 if $v_j$ is in the neighbor set of $v_k$ and 0 otherwise, and $\rho^t$ denote the value of $\rho$ at iteration $t$ during the EM process. Next we analyze the two procedures of wedge closing or random insertion given a starting node (i.e., first selected node) $v_i$.  We can calculate the probabilities based on the following: (1) for the random insertions with probability $(1-\rho)$ and selecting $v_j$ as the second node with probability $\bm{\pi}_i$; (2) the wedge closing with probability $\rho$ and the probability we were able to perform a two-hop walk to $v_j$ is based on first having $v_k$ that is a mutual neighbor of $v_i$ and $v_j$ (i.e., $v_k \in N_i$ and $v_k \in N_j$) and then the walk continues to $v_j$ (i.e., selecting $v_j$ from the $d_k$ neighbors of $v_k$) once arriving at the mutual neighbor $v_k$. Therefore, we can formulate the conditional probabilities for placing the edge $e_{ij}$ given $\rho$, the starting node $v_i$ and the method for either the random insertion or wedge closing procedure (that is represented with $z_{ij}$) are as follows, respectively:
\begin{align}
&P(e_{ij}|z_{ij}=0,v_i,\rho^t) = (1-\rho^t) (\bm{\pi}_i) \nonumber \\
&P(e_{ij}|z_{ij}=1,v_i,\rho^t) = \rho^t \sum\limits_{v_k \in N_i} \Big(\frac{\mathbb{I}[v_j \in N_k]}{d_i}\Big) \Big( \frac{1}{d_k} \Big) \nonumber
\end{align}

For the calculation of the expectation of $z_{ij}$ given $\rho^t$, $e_{ij}$, and the starting node $v_i$, the conditional probability of $z_{ij}$ can be defined using the Bayes' Rule as follows:
\begin{align}
P(z_{ij}=1|e_{ij},v_i,\rho^t) = \frac{P(e_{ij}|z_{ij}=1,v_i,\rho^t)}{P(e_{ij}|z_{ij}=1,v_i,\rho^t) + P(e_{ij}|z_{ij}=0,v_i,\rho^t)}
\end{align}

\noindent which calculates the probability of $z_{ij}$ being 1 based on the probability of the edge being created by wedge closure over the probability the edge $e_{ij}$ is expected  to get created. This leads to the expectation of $z_{ij}$ to $\mathbb{E}[z_{ij}|\rho^t] = P(z_{ij}=1|e_{ij},v_i,\rho^t)$. Furthermore, the maximization for the expectation can be calculated via sampling a set of edges $\mathcal{S}$ uniformly from $\mathcal{E}$. Then, due to the fact $z_{ij}$ is conditionally independent, we can individually calculate the expectation of $z_{ij}$ for each edge in $\mathcal{S}$ and then take the average across the set of edges sampled as:
\begin{align}
p^{t+1} = \frac{1}{\mathcal{S}} \sum\limits_{e_{ij} \in \mathcal{S}} \mathbb{E}[z_{ij}|\rho^t]
\end{align}

\vspace{-2ex}
\subsubsection{Learning $\beta$}  

Note that we have calculated $\Delta_B$ from the input network that denotes the percentage of triangles that were adhering to balance theory. We seek to approximate the expected number of triangles BSCL will construct through the wedge closure and the random edge insertion methods on average for each edge added to the network. Let us denote the values we calculate for these two methods as $\Delta_\mathcal{G}^{triangle}$ and $\Delta_\mathcal{G}^{random}$, respectively, which will be calculated with respect to both $\rho$ and $\alpha$. Furthermore, we will calculate what percent of these we expect to be balanced as $\Delta_{\mathcal{G}B}^{triangle}$ and  $\Delta_{\mathcal{G}B}^{random}$. Details of estimating $\Delta_\mathcal{G}^{triangle}$, $\Delta_\mathcal{G}^{random}$, and $\Delta_{\mathcal{G}B}^{random}$ will be discussed later.  To correctly maintain the percentage of triangles being balanced in the synthetic network, we desire the following:
\begin{align}
\Delta_B = \frac{\Delta_{\mathcal{G}B}^{triangle} + \Delta_{\mathcal{G}B}^{random}}{\Delta_\mathcal{G}^{triangle} + \Delta_\mathcal{G}^{random}} \nonumber
\end{align}
which simply states the combined balanced percentage from the two methods should be the balanced percentage of the input network. Then, we can calculate the above mentioned values and if we let $\Delta_{\mathcal{G}B}^{triangle} = \beta \Delta_{\mathcal{G}}^{triangle}$, which denotes that $\beta$ percent of the triangles we close via the wedge closing procedure are balanced, then we can solve $\beta$ and obtain the below:
\begin{align}
\beta = \frac{\Delta_B \big(\Delta_{\mathcal{G}}^{triangle} + \Delta_{\mathcal{G}}^{random}\big) - \Delta_{\mathcal{G}B}^{random}}{\Delta_{\mathcal{G}}^{triangle}}\label{eq:update_beta}
\end{align}
Next,  we discuss how to estimate $\Delta_\mathcal{G}^{random}$, $\Delta_{\mathcal{G}B}^{random}$ and $\Delta_\mathcal{G}^{triangle}$. 

{\bf Estimating $\Delta_\mathcal{G}^{random}$}: We note that the starting set of edges are constructed with the FCL method and edge signs randomly assigned to them. Furthermore, each edge will have been added into the network with probability approximately equal to $p_{ij} = \frac{d_i d_j}{2M}$. We note that the expected number of common neighbors between two vertices $v_i$ and $v_j$ would be equivalent to the number of triangles that get created if the edge $e_{ij}$ was inserted into the network where we denote this number of triangles to be $\Delta_{ij}^{random}$.

To obtain the number of common neighbors for $v_i$ and $v_j$, we calculate the probability that $v_l \in V_I \backslash \{v_i,v_j\}$ is a common neighbor based on the probability there exists an edge from $v_l$ to both $v_i$ and $v_j$. Note that after having the probability of the existence for the first edge $e_{il}$, we must subtract 1 from $d_l$, since we have already conditioned on the existence of the first edge $e_{il}$, thus causing $v_l$ to have one less opportunity to connect to $v_j$. We formulate this idea as the following:
\vspace{-2ex}
\begin{align}
\Delta_{ij}^{random} & = \sum\limits_{v_l \in V \backslash \{v_i,v_j\} }^{}{ \Big( \frac{d_i d_l}{2M} \Big) \Big( \frac{d_j(d_l - 1)}{2M} \Big) } \nonumber \\
& = \Big( \frac{d_i d_j}{2M} \Big)  \sum\limits_{v_l \in V \backslash \{v_i,v_j\}}^{}{ \Big( \frac{d_l (d_l - 1)}{2M} \Big) } \nonumber 
\end{align}
Next we present the average value of $\Delta_{ij}^{random}$ across all possible unordered pairs of vertices as follows:
\begin{align} 
\Delta_{\mathcal{G}}^{random} & = \frac{1}{\frac{1}{2}N(N-1)} \sum\limits_{i=1}^{N-1} \sum\limits_{j=i+1}^{N} \Delta_{ij}^{random} \nonumber 
\end{align} 
\noindent where $\Delta_{\mathcal{G}}^{random}$ is used to denote the average triangles constructed by a randomly inserted edge in the model. % in the CL model. 
We note that the above would require $O(N^3)$ time to compute, but using the fact that $2M=N*avg(d)$, we can use the following approximation if we treat the summation such that it includes $v_i$ and $v_j$ instead of excluding them. We use $avg(d)$ to denotes the average degree and $avg(d^2)$ represents the average value of squared degrees. Then we have: 
\begin{align}
\sum\limits_{v_l \in V \backslash \{v_i,v_j\}}^{}{ \Big( \frac{d_l (d_l - 1)}{2M} \Big) } &\approx  \sum
\limits_{v_l \in V}^{}{ \Big( \frac{d_l (d_l - 1)}{2M} \Big) } \nonumber \\
& = \frac{(d_1^2 + d_2^2 + \dots + d_N^2) - (d_1 + \dots + d_N)}{2M} \nonumber \\
& = \frac{\big(avg(d^2)N\big) - \big(avg(d)N\big)}{N*avg(d)} \nonumber \\
& = \Big( \frac{avg(d^2) - avg(d)}{avg(d)} \Big) \label{eq:assumption_simplification}
\end{align} 
\noindent We therefore can rewrite $\Delta_{\mathcal{G}}^{random}$ as follows:
\begin{align} 
\Delta_{\mathcal{G}}^{random} &\approx \frac{avg(d^2) - avg(d)}{avg(d)MN(N-1)} \sum\limits_{i=1}^{N-1} \sum\limits_{j=i+1}^{N} d_i d_j \nonumber
\end{align}

First we note that we only need to compute $avg(d)$ and $avg(d^2)$ once (which can be performed in $O(N)$ time). Second, for $\sum_{i=1}^{N-1}$ $d_i \sum_{j=i+1}^{N} d_j $, rather than iterating over the nested sum of $j=i+1$ to $N$, we can instead use dynamic programming to construct a vector ${\bf s}$ where $s_i$ represents $\sum_{j=i+1}^{N} d_j$. We can construct this vector ${\bf s}$ starting with $s_i = d_N$ when $i=N-1$ and then recursively filling in the vector using $s_{i-1} = s_{i} + d_{i}$, which can be performed in linear time in relation to the number of vertices $N$. Therefore the below approximation for $\Delta_{\mathcal{G}}^{random}$ can be performed in $O(N)$ time instead of $O(N^3)$. 
\vspace{-1ex}
\begin{align} 
\Delta_{\mathcal{G}}^{random} &\approx \frac{avg(d^2) - avg(d)}{avg(d)MN(N-1)} \sum\limits_{i=1}^{N-1} d_i  s_i \label{eq:delta_random_count}
\end{align}

{\bf Estimating $\Delta_{\mathcal{G}B}^{random}$}: We further analyze beyond our calculation of $\Delta_{\mathcal{G}}^{random}$ by examining the wedges (i.e., common neighbors) to be one of the following formations: $(\{+,+\},$ $\{+,-\},$ $\{-,+\},$ $\{-,-\})$, where $\{+,-\}$ is used to represent the wedge formed by edges $e_{il}$ and $e_{lj}$ and their signs are positive and negative, respectively. Note that we first initialize our model with edges from FCL and select edge signs to perfectly match the sign distribution of the original network, and then we attempt to correctly maintain this distribution with the parameter $\alpha$; hence we assume that all wedges were created by the original sign distribution (where $\eta$ is the probability of a link being positive). Below, we use $\Delta_{ij}^{random+-}$ to represent the number of wedges that would be closed into triangles when adding the edge $e_{ij}$ and were formed with a wedge of type $\{+,-\}$. The definitions for all the wedge types are: 
\begin{align}
&\Delta_{ij}^{random++} = \eta \eta \Delta_{ij}^{random}  \nonumber \\
&\Delta_{ij}^{random+-} = \Delta_{ij}^{random-+} = \eta (1-\eta) \Delta_{ij}^{random}   \nonumber \\
&\Delta_{ij}^{random--} = (1-\eta) (1-\eta)\Delta_{ij}^{random}   \nonumber 
\end{align}

The expected number of balanced triangles that would be created if the edge $e_{ij}$ is inserted randomly, $\Delta^{random}_{ijB}$, can be obtained via the expected number of wedges of different types and the corrected positive link probability $\alpha$ as: 
\begin{align}
\Delta_{ijB}^{random}  = \alpha\Delta_{ij}^{random++} + (1-\alpha)\Delta_{ij}^{random+-} \nonumber \\+ (1-\alpha)\Delta_{ij}^{random-+} + \alpha\Delta_{ij}^{random--} \label{eq:delta_random_balance} 
\end{align}
\noindent where for a wedge with two existing edges to close to a balanced triangle, the added third edge would need to have a sign such that there are an even number of negative links in the resulting triangle, according to balance theory. This can also be extended for the calculation of $\Delta_{\mathcal{G}B}^{random}$ by averaging across all edges.

{\bf Estimating $\Delta_\mathcal{G}^{triangle}$}: Similarly, we will calculate the expected total number of triangles and the balanced percentage when using the wedge closing procedure. The main idea for this wedge closure is that we are guaranteed to select vertices such that we have at least one triangle being created each time. We then need to also add the expected number of triangles that would be created randomly by other common neighbors of $v_i$ and $v_j$ (similar to the random edge insertion case above). Note that we must however discount the degree of $v_i$ and $v_j$ by 1, since in this method, we have already explicitly used one of the links coming from both to discover this one common neighbor for the wedge closure edge insertion. Let us denote the selected common neighbor as $v_c$, which forms a wedge with the edges $e_{ic}$ and $e_{cj}$.
\begin{align}
\Delta_{ij}^{triangle} &= 1 + \sum\limits_{v_l \in V \backslash \{v_i,v_j,v_c\} }^{}{ \Big( \frac{(d_i - 1) d_l}{2M} \Big) \Big( \frac{(d_j - 1)(d_l - 1)}{2M} \Big) } \nonumber \\
& = 1 + \Big( \frac{d_i d_j - d_i - d_j + 1}{2M} \Big)  \sum\limits_{v_l \in V \backslash \{v_i,v_j,v_c\}}^{}{ \Big( \frac{d_l (d_l - 1)}{2M} \Big) } \nonumber 
\end{align}

We can make a similar approximation as Eq.~(\ref{eq:assumption_simplification}), and can simplify the formulation of $\Delta_{ij}^{triangle}$ as:
\begin{align}
\Delta_{ij}^{triangle} \approx 1 + \Big( \frac{d_i d_j - d_i - d_j + 1}{2M} \Big) \Big( \frac{avg(d^2) - avg(d)}{avg(d)} \Big) \nonumber
\end{align}

Then we can also calculate $\Delta_{\mathcal{G}}^{triangle}$ similar to $\Delta_{\mathcal{G}}^{random}$, which results in the following:
\begin{align} 
\Delta_{\mathcal{G}}^{triangle} &\approx 1 + \frac{avg(d^2) - avg(d)}{avg(d)MN(N-1)} \sum\limits_{i=1}^{N-1} \big( (d_i - 1)(s_i -N + i) \big) \nonumber 
\end{align}

\subsubsection{Learning $\bm{\alpha}$}

Here we want to estimate the percentage of positive edges. Therefore, we will examine this percentage for both the wedge closing and the random edge insertion, which are denoted as $\eta^{triangle}$ and $\eta^{random}$, respectively. Then, if we have these two values, we can correctly maintain the percentage of positive links in the synthetic network as the following:
\begin{align}
\eta = \rho \eta^{triangle} + (1-\rho) \eta^{random} \label{eq:solve_eta}
\end{align}
The above is due to the fact that the wedge closing procedure will insert $\rho$ percent of the links into the generated network while the random insertion method will construct $(1-\rho)$ percent.  

For the wedge closure, we examine the probability of the four types of wedges: \{+,+\}, \{+,-\}, \{-,+\} and \{-,-\}. Then we define their probabilities of existing in the network to be P(\{+,+\}), P(\{+,-\}), P(\{-,+\}) and P(\{-,-\}). Next we note that the wedge \{+,+\} and \{-,-\} would result in a positive edge being created with probability $\beta$, while wedge \{+,-\} and \{-,+\} would only provide a positive edge with probability $(1-\beta)$. The reason for this is due to the fact that balance theory would be controlling the third edge sign and while the first two require a positive link to adhere to balance theory; while it would be when we construct more unbalanced triangles that the later two wedge types would result in a positive link insertion. Therefore, we denote the probability of inserting a positive edge with the wedge closure to be the following:
\begin{align}
\eta^{triangle} = ~&\beta \Big( P(\{+,+\}) + P(\{-,-\}) \Big) \nonumber \\
	&+ (1-\beta) \Big( P(\{+,-\}) + P(\{-,+\}) \Big) \label{eq:eta_triangle}
\end{align}
If we assume that $\alpha$ and $\beta$ are correctly solved, then the expected probability for the wedges is based on $\eta$ and $(1-\eta)$. We can therefore rewrite Eq.~\ref{eq:eta_triangle} as:
\begin{align}
\eta^{triangle} = &\beta \Big( (\eta \eta) + (1-\eta)(1-\eta) \Big) \nonumber \\
	& + (1-\beta) \Big( \eta(1-\eta) + (1-\eta)\eta \Big) \label{eq:eta_triangle2}
\end{align}

Next, we show how to calculate $\eta^{random}$. We notice that $\alpha$ will be the percentage of links we construct to be positive and this process will happen $(1-\rho)$ percent of the time. Thus, $\eta^{random} = (1-\rho) \alpha$. We can now substitute $\eta^{triangle}$ and $\eta^{random}$ into Eq.~\ref{eq:solve_eta} and solve for $\alpha$ as follows:
\begin{align}
\alpha = \frac{1}{(1-\rho)} \Bigg[ \eta - \rho \Bigg(& \beta \Big( (\eta\eta) + (1-\eta)(1-\eta) \Big) \nonumber \\
&+ (1-\beta)\Big( \eta(1-\eta) + (1-\eta)\eta \Big) \Bigg) \Bigg] \label{eq:alpha_update}
\end{align}

\vspace{-1ex}
\subsection{Time Complexity}
We first discuss the running time of the learning algorithm and then the time needed for the network generation process. The preprocessing needed for the learning algorithm is to determine the probability of edges being positive, $\eta$, and the probability of triangles in the network being balanced and adhering to balance theory, $\Delta_B$, of the original network. We can determine $\eta$ trivially in $O(M)$. However, $\Delta_B$ can be reduced the complexity of triangle listing algorithms, which can be easily performed using classical methods in $O(min(M^{3/2},Md_{max}))$, where $d_{max}$ is the maximum degree in the network\cite{schank2005finding}.  
The learning process for the parameter $\rho$ is $O(sI)$ where $I$ is the number of iterations in the EM method and $s=|\mathcal{S}|$ is the number of edges sampled for each iteration. The running time of $\beta$ and $\alpha$ can be determined as follows. We initially calculate the expected number of triangles added by each of the processes of BSCL, which takes $O(N)$ instead of $O(N^3)$ due to the approximation used and dynamic programming approach. 
Then the update equations (i.e., Eqs. (\ref{eq:update_beta}) and (\ref{eq:alpha_update})) can both be performed in $O(1)$ time. Thus when allowing for $I'$ maximum iterations of the alternating update process between $\alpha$ and $\beta$ (which empirically only takes a small constant number of iterations to converge), we have the overall learning time complexity for BSCL as $O(min(m^{3/2},md_{max}) + N^2   +   sI  +  I')$. The generation process of BSCL, is built upon the fact that the running time for TCL is shown to be $O(N+M)$ in \cite{pfeiffer2012fast}. The triangle closing process of determining the best sign selection based on the set of triangles being closed, is reduced to the complexity of common neighbors between two vertices, which is known to be $O(d_{max}^2)$. 
Thus the generation process of BSCL is $O(N + Md_{max}^2)$.

\section{Experiments}
In this section, we conduct experiments to evaluate the effectiveness of the proposed signed network model. In particular, we try to answer two questions via our experiments - (1) can the proposed model, BSCL, effectively maintain signed network properties? and (2) is the parameter learning algorithm able to learn the appropriate parameter values from the input signed network? We first introduce the datasets, then design experiments to seek answers for the above two questions by first comparing BSCL against baseline models, and finally we perform an analysis of the parameter learning algorithm.

\subsection{Datasets}
For our study of signed network modeling, we collect three signed network datasets, i.e., Bitcoin-Alpha, Bitcoin-OTC, and Epinions. We provide more details of the datasets in Table~\ref{tab:datasets}. 

\begin{table}
\begin{center}
\caption{Statistics of three signed social networks.}
\label{tab:datasets}
\begin{tabular}{c|c|c|c}
	\hline
	Network 			& 	N 		& (E, $\eta$) 	& $\Delta_B$ \\\hline
	Bitcoin-Alpha  	& 3,784 		&	(14,145  , 0.915)			&	0.862		\\	
	Bitcoin-OTC 		& 5,901 		&	(21,522  , 0.867)			&	0.869		\\	
	Epinions  		& 131,580 	&	(711,210 , 0.830)			&	0.892		\\ \hline
\end{tabular}
%\vspace{-0.15in}
\end{center}
\end{table}

The Bitcoin-Alpha and Bitcoin-OTC datasets were collected from Bitcoin Alpha\footnote{http://www.btcalpha.com} and Bitcoin OTC\footnote{http://www.bitcoin-otc.com}, respectively~\cite{kumar2016edge}. In these sites, users can buy and sell things in an open marketplace using Bitcoins. For the safety of the users, they form online trust networks to protect against scammers. We also have collected a dataset from Epinions\footnote{http://www.epinions.com}, which is a product review site where users can mark their trust or distrust to other users, representing positive and negative links in the signed network, respectively.  Note that, since we focus on undirected and unweighted signed networks, we ignore the directions of signed links in these datasets. 

  \begin{table*}[t]
    \begin{minipage}{.44\textwidth}
      \centering
     \caption{\label{tab:sign_ratio} Positive/Negative Link Sign Distribution.}
             \begin{tabular}{|c|c|c|c|c|c|c|c|c|}	\hline
Links Positive 	&	Real		&	Ants14	&	Evo07	&	BSCL	\\ \hline
Bitcoin-Alpha		&	0.915		&	0.741	&	0.917	&	0.912	\\ 
Bitcoin-OTC		&	0.867		&	0.740	&	0.869	&	0.860	\\ 
Epinions			&	0.830		&	0.930	&	0.830	&	0.808	\\
 \hline
Absolute Difference &		&	0.401	&	0.004	&	0.031 \\ \hline
\end{tabular}
    \end{minipage}
    \begin{minipage}{.55\textwidth}
      \centering
   \caption{\label{tab:balance_table} Proportion of Triangles Balanced.}
             \begin{tabular}{|c|c|c|c|c|c|c|c|c|}	\hline
Percent Balanced		&	Real		&	STCL/SKron	&	Ants14	&	Evo07	&	BSCL		\\ \hline
Bitcoin-Alpha			&	0.862		&	0.786			&	0.787	&	0.750	&	0.802	\\
Bitcoin-OTC				&	0.869		&	0.698			&	0.817	&	0.677	&	0.752	\\ 
Epinions				&	0.892		&	0.644			&	0.939	&	0.639	&	0.748	\\ \hline
Absolute Difference & 					&	0.484			&	0.174	&	0.557	&	0.321 \\ \hline
\end{tabular}
    \end{minipage}
  \end{table*}
  
\begin{table}[t]
             \begin{center}
             \small
             \vskip -1em
             \caption{\label{tab:triangle_type_dist} Distribution of Triangle Types (Bitcoin-OTC).}
             \begin{tabular}{|c|c|c|c|c|c|c|c|c|}	\hline
Triad Type		&	Real	&	STCL/SKron&	Ants14	&	Evo07	&	BSCL	\\ \hline
\{+,+,+\}			&	0.724	&	0.652		&	0.445	&	0.613	&	0.707	\\ 
\{+,+,-\}			&	0.122	&	0.300		&	0.170	&	0.323	&	0.246	\\ 
\{+,-,-\}			&	0.145	&	0.046		&	0.372	&	0.064	&	0.045	\\ 
\{-,-,-\}			&	0.009	&	0.002		&	0.013	&	0.000	&	0.002	\\ \hline
Absolute Difference & 		&	0.356		&	0.558	&	0.401	&	0.248 \\ \hline
\end{tabular}
               \vspace{-0.09in}
               \end{center}
\end{table}  

\subsection{Network Generation Experiments}
The first set of experiments are to compare the network properties of the resulting generated networks from our model and the baselines. These properties will be used as a metric to determine how well the models are able to capture the underlying dynamics of signed networks. More specifically, we will focus on the three key signed network properties - (1) degree distribution; (2) positive/negative link ratio and (3) proportion of balance/unbalanced triangles suggested by balance theory. Note that we also present the local clustering coefficient distribution and the triangle distribution (in relation to the edge signs in the triangles). Our results are the averaged results of 10 generated networks for each of the methods on each dataset.

The first group of two baselines are existing signed network models: (1) \textbf{Ants14:} it is an interaction-based model for signed networks based on using ants to lay pheromone on edges \cite{vukavsinovic2014modeling}; and (2)  \textbf{Evo07:} it is an evolutionary model for signed networks that had a ``friendliness'' index that controls the probability of positive or negative links and also a parameter that controls the maximum amount of unbalance~\cite{ludwig2007evolutionary}. Note that for Ants14, we perform a grid search on the parameter space for its 6 parameters according to the values reported in \cite{vukavsinovic2014modeling}. Similarly, for Evo07, a grid search was performed for the two parameters. 

\begin{figure}[t]
\begin{subfigure}{0.22\textwidth}
  \includegraphics[height=3cm]{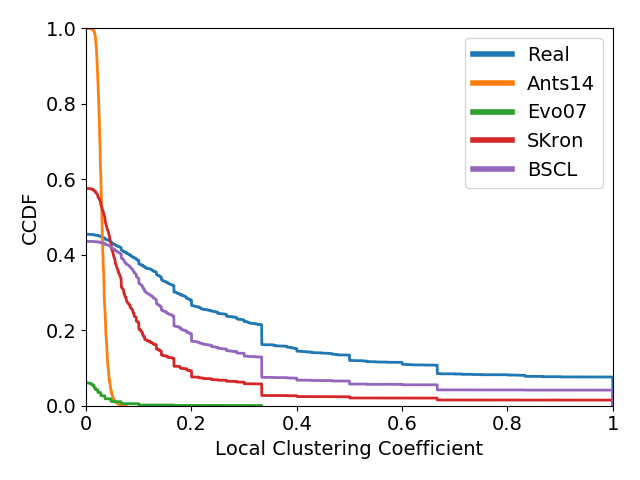}
\caption{Bitcoin-Alpha Local Clustering} \label{fig:bitcoin_alpha_lcc}
\end{subfigure}
\hspace*{\fill} % separation between the subfigures
\begin{subfigure}{0.22\textwidth}
  \includegraphics[height=3cm]{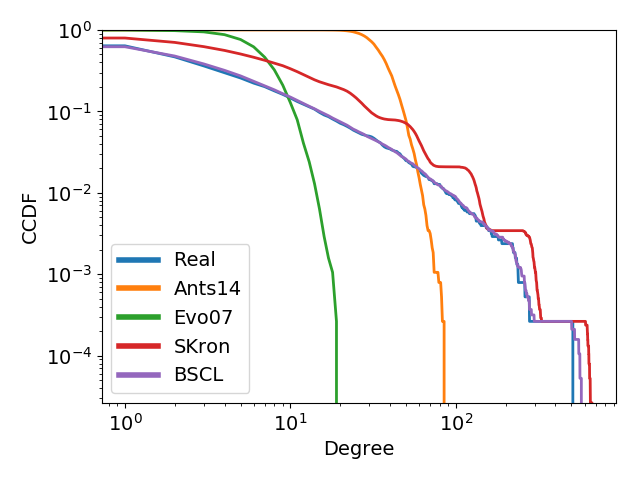}
\caption{Bitcoin-Alpha Degree Dist.} \label{fig:bitcoin_alpha_dd}
\end{subfigure}\\
\begin{subfigure}{0.22\textwidth}
  \includegraphics[height=3cm]{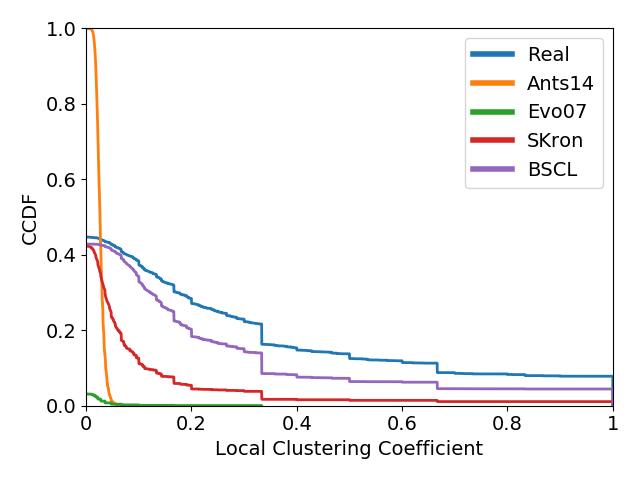}
\caption{Bitcoin-OTC Local Clustering} \label{fig:bitcoin_otc_lcc}
\end{subfigure}
\hspace*{\fill} % separation between the subfigures
\begin{subfigure}{0.22\textwidth}
  \includegraphics[height=3cm]{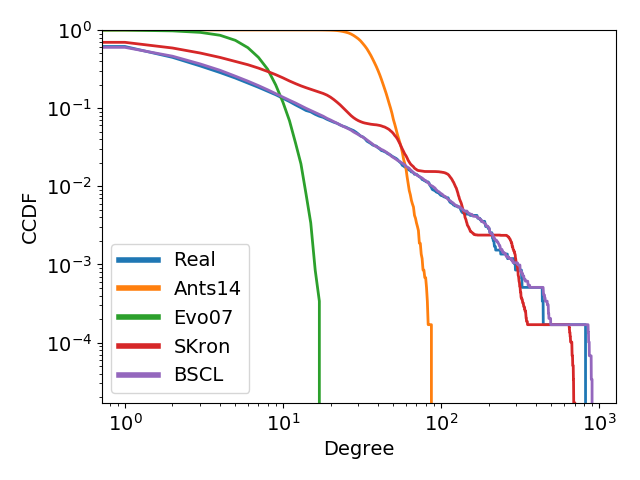}
\caption{Bitcoin-OTC Degree Dist.} \label{fig:bitcoin_otc_dd}
\end{subfigure}\\
\begin{subfigure}{0.22\textwidth}
  \includegraphics[height=3cm]{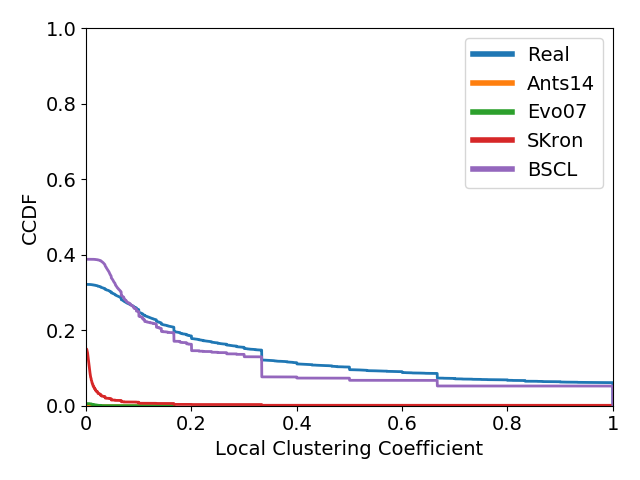}
\caption{Epinions Local Clustering} \label{fig:epinions_lcc}
\end{subfigure}
\hspace*{\fill} % separation between the subfigures
\begin{subfigure}{0.22\textwidth}
  \includegraphics[height=3cm]{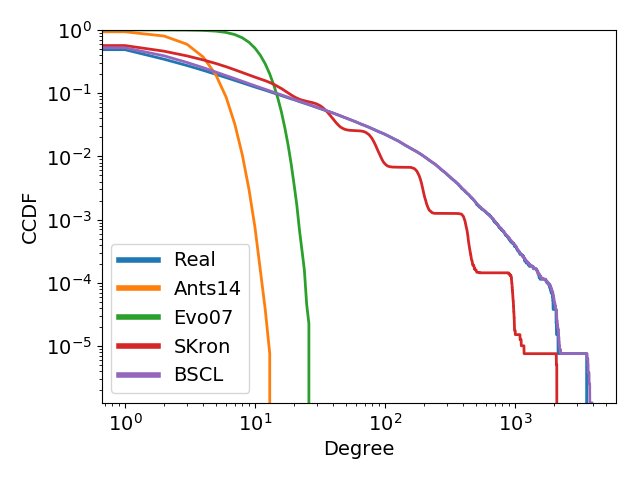}
\caption{Epinions Degree Dist.} \label{fig:epinions_dd}
\end{subfigure}
\caption{Degree Distribution and Local Clustering Coefficient.} \label{fig:dd_lcc}
\end{figure}

The next two baselines are built upon two popular unsigned generative models. We first convert the network to unsigned by ignoring the links, run the baseline model, and then randomly assign signs to the edges such that the global sign distribution is maintained using $\eta$. They are -- (1) \textbf{STCL} from the TCL model \cite{pfeiffer2012fast}; and (2) \textbf{SKron} from the Kronecker Product model \cite{leskovec2010kronecker}. 

The results of the properties that are in common with unsigned networks (i.e., the degree distribution and the local clustering coefficient) can be can be seen in Figure~\ref{fig:dd_lcc}. Note that we do not show the STCL model in Figure~\ref{fig:dd_lcc} since it performs near identically on the degree distribution to BSCL as they are both based on the Transitive Chung-Lu model and therefore can very closely approximate the degree distribution. However, it can be seen that the two signed network baselines, Ants14 and Evo07, perform very poorly and do not even appear to follow a power-law distribution. We mention that SKron is not able to exactly model the degree distribution, but does not perform as poorly as the two existing signed network baselines. For the two existing signed network models similar poor findings can be found for the local clustering coefficient. Our proposed model BSCL along with STCL perform the best. The SKron model has some clustering, but not near that of the original input network. 

In Table \ref{tab:sign_ratio}, we show the positive/negative link ratio; while in Table \ref{tab:balance_table}, we present the proportion balance/unbalance triangles. We make a comparison to the two existing signed network methods, and also then follow with a comparison of BSCL to STCL and SKron which are the two modified unsigned network models.

In Bitcoin-Alpha dataset, our model BSCL is able to achieve the closest proportion of balance triangles. Then in the Bitcoin-OTC dataset, the Ants14 performs the best in terms of the proportion of balance in the network. We further show a fine-grained comparison by separating the four types of triads on the Bitcoin-OTC dataset in Table \ref{tab:triangle_type_dist}. We notice that Ants14 achieves this by drastically changing the distribution among the four triangle types. However, our model performs the best overall in terms of the triangle distribution. Similarly we notice a drastically low overall clustering in the Ants14 output networks as seen in the local clustering coefficient plot in Figure~\ref{fig:bitcoin_otc_lcc}. Furthermore, although the Ants14 method more closely resembles the percentage of triangles being balanced across the three signed networks (having a smaller absolute difference than BSCL), we can observe that this comes at a trade-off of having a very inconsistent percentage of links being positive in the network as compared to the input signed network, which can be seen in the absolute difference row of Table~\ref{tab:sign_ratio}. Note that in both of the Bitcoin datasets, our model BSCL is able to achieve better performance than the baselines in terms of the triangle distributions, while only at the expense of sacrificing $<1\%$ in terms of matching the correct positive link percentage, $\eta$, of the input networks. The results are similar in the Epinions dataset.

When comparing BSCL with STCL and SKron we mention that by design these other two models will always have the exact percentage of positive links and their expected triangle distribution can be calculated as: $\big( \eta \eta \eta \big)$, $\big( 3\times(\eta\eta(1-\eta)) \big)$, $\big( 3\times(\eta(1-\eta)(1-\eta)) \big)$, and $\big( (1-\eta)(1-\eta)(1-\eta) \big)$ for the \{+,+,+\}, \{+,+,-\}, \{+,-,-\}, and \{-,-,-\} triangle types, respectively. We can observe that in terms of the absolute difference between the percentages in each of these triangle types, the BSCL method performs much better having an absolute difference of only 0.248 in the Bitcoin-OTC dataset while the straight-forward modification to the unsigned network models of maintaining the positive link percentage correctly results in an absolute difference of 0.356. Similarly, when looking at the percentage of triangles adhering to balance theory across the three signed network datasets BSCL has a value of only 0.321 while STCL and SKron share an absolute difference of 0.484. 

Overall we can see that the Ants14 model favors capturing the percentage of triangles adhering to balance theory, but at the sacrifice of the triangle distribution (in regards to the triangle edge signs) and performing just as poorly at maintaining the positive/negative link ratio. On the other hand, the Evo07 model is able to correctly maintain the positive/negative link ratio, but struggles to maintain reasonable percentage of balanced triangles when examining across the three datasets. Our model BSCL overall outperforms the two previous signed network models as seen in the Tables and Figures. Furthermore, we can see BSCL out performs STCL and SKron in terms of maintaining the percentage of balanced triangles and triangle distribution. We note that BSCL finds this improvement in other signed network properties while only losing about ~1\% on average across each the three datasets in maintaining the positive/negative link ratio.   

\subsection{Parameter Learning Experiments} \label{subsec:param_learning_exp}

The second set of experiments are designed to test the learning algorithm we have proposed in determining appropriate parameters for BSCL.
Here we first utilized natural and intuitive heuristics for setting the parameter values of $\alpha$ and $\beta$ to further evaluate the effectiveness of our parameter learning algorithms. More specifically, the value of $\eta$ (i.e., the real network's percentage of positive links) is used as a natural choice for the value of $\alpha$ (i.e., the percentage of the time a randomly inserted edge is positive) when not utilizing the proposed learning algorithm described in Section~\ref{subsec:param_learning} for BSCL. Similarly, the value $\Delta_B$ (i.e., the real network's percentage of triangles that are balanced) can be chosen as the value for $\beta$ (i.e., the percentage of time we explicitly close a wedge into a balanced triangle) if we were to not use our proposed learning algorithm.  Table~\ref{tab:alpha_beta_absolute_difference} contains the performance comparison (in relation to the link sign distribution, proportion of triangles balanced, and distribution of triangle types) between the proposed parameter learning algorithms selected values against the above mentioned heuristically picked values for $\alpha$ and $\beta$. Note that these absolute difference values are averaged across the three real-world signed networks. We can observe that in all three properties our parameter learning algorithm significantly out performs the most natural heuristically picked values and thus providing further evidence our model can learn parameters to more accurately generate synthetic variants of the real input signed network.

\begin{table}
\small
\begin{center}
\caption{Absolute difference from the generated networks to the real signed networks averaged over the three datasets for each respective property.}
\label{tab:alpha_beta_absolute_difference}
\begin{tabular}{|c|c|c|}
	\hline
	 	& 	\begin{tabular}{c} BSCL\\ (learning $\alpha$ \& $\beta$) \end{tabular} & \begin{tabular}{c} BSCL\\ ($\alpha = \eta$ \& $\beta = \Delta_B$) \end{tabular} \\ \hline 
                \begin{tabular}{c} Sign \\ Distribution \end{tabular} & 0.031 & 0.040 \\ \hline
        \begin{tabular}{c} Proportion \\ Balanced Triangles \end{tabular} & 0.107 & 0.164 \\ \hline 
        \begin{tabular}{c} Distribution \\ Triangle Types \end{tabular} & 0.220 & 0.351 \\ \hline
\end{tabular}
\end{center}
\end{table}

\begin{figure}[]
\begin{subfigure}{0.48\linewidth}
  \includegraphics[width=\linewidth]{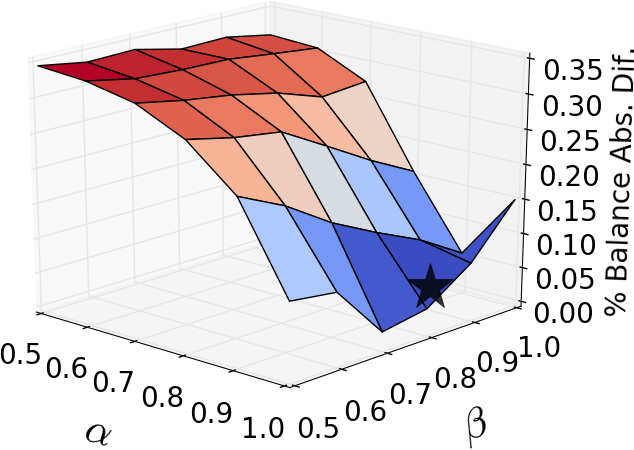}
\caption{Bitcoin-Alpha\\ \% Balanced Triangles} \label{fig:param_analysis_bitcoin_alpha_balance}
\end{subfigure}
\hspace*{\fill} % separation between the subfigures
\begin{subfigure}{0.48\linewidth}
  \includegraphics[width=\linewidth]{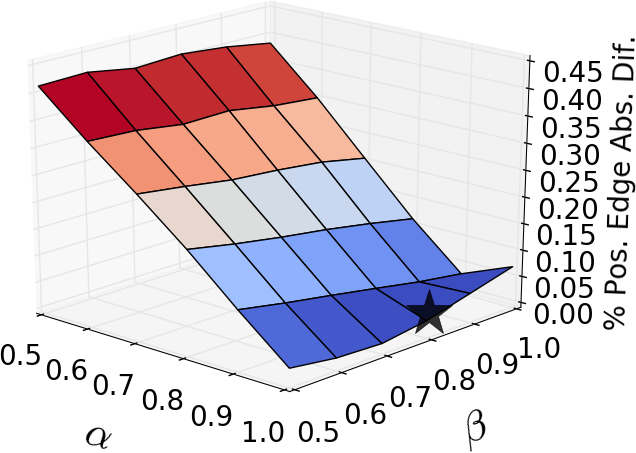}
\caption{Bitcoin-Alpha \\Positve/Negative Ratio} \label{fig:param_analysis_bitcoin_alpha_posneg}
\end{subfigure}
\caption{Parameter Learning Analysis.} \label{fig:param_analysis}
\vspace{-3ex}
\end{figure}

For a more detailed analysis we also perform a grid search across a reasonable area of the parameter space for $\alpha$ and $\beta$ to obtain optimal parameters. Then we compare the performance of the learnt parameters and the searched optimal parameters to demonstrate the ability of the proposed parameter learning algorithm. We only present the results in Figures \ref{fig:param_analysis} in terms of percentage of balanced triangles and positive/negative link ratio for the Bitcion Alpha dataset, since we have similar observations for Bitcion OTC and Epinions with other settings.  Note that the z-axis is the absolute difference away from the true input networks value (where lower is better). The ``stars'' in the figures are the coordinates along the x- and y-axis for the learned parameter. From the figures, it looks convincing that indeed our parameter learning algorithm is able to find appropriate parameters for the input network.

\section{Related Work}
In this section, we briefly review the work that have attempted to model signed networks. However, there is no existing work focused on signed network generative modeling. This means they do not provide a mechanism for learning parameters to construct similar signed networks to a given input network, but instead parameters would need to be hand-picked.

In \cite{ludwig2007evolutionary} Ludwig and Abell proposed an evolutionary model for signed networks that made use of two parameters. One of the parameters is defined as the ``friendliness'' index and controls the probability a randomly created edge will be positive or negative. The other parameter is a threshold that represents the maximum amount of unbalanced triangles each user of the network can participate in. 
The primary downside of this method is that it is not a generative model (i.e., parameters must be hand-picked and cannot be learned). A more recent model was based on a game-theoretic approach to model the formation of signed networks \cite{malekzadeh2011social}. In this method each user tries to minimize their social stress by attempting to individually follow balance theory to maximize the number of balanced minus unbalanced triangles they are participating in. This work is different in that they focused more on the theoretical analysis of balance theory in signed networks and have the objective of taking an existing network and optimizing the edges by either removing, flipping signs, or including new edges, such that the network becomes balanced (i.e., all triangles in the network are balanced). This work is not comparable to ours in that they do not attempt to generate synthetic networks with similar properties, but instead focus on adjusting/evolving existing networks to build complete networks that adhere to balance theory. In \cite{vukavsinovic2014modeling}, an interaction-based model is presented to construct signed networks with a focus on preserving balanced triangles in the network. Their model uses an ant-based algorithm where pheromone is placed on edges where a positive (negative) amount represents a positive(negative) edge with the magnitude of the pheromone being the strength of the edge. During each step of the algorithm, for each node, a local and global process is performed. The global step allows for longer distance connections to be made proportional to the degree of the nodes. In comparison, the local step is used for reinforcing the pheromone on a triangle in the network and adjusts the third edge to adhere to balance theory based on the first two edges in the triangle. Note that this method contains 6 parameters and none can be learned by their model, but instead, require them to be hand-crafted.

\section{Conclusion}
 Very few works have focused modeling of signed networks, as most have focused on unsigned networks. Generative network models specifically have been shown to provide a vast amount of insight into the underlying network structures. In signed networks specifically, social theories have been developed to describe the dynamics and mechanisms that drive the structural appearance of signed networks. Structural balance theory is one such social theory. We have proposed our Balanced Signed Chung-Lu model (BSCL) with the objective of preserving three key properties of signed networks - (1) degree distribution; (2) positive/negative link ratio; and (3) proportion of balance/unbalanced triangles suggested by balance theory. To achieve this, we introduced a triangle balancing parameter and a sign balancing parameter to control the distribution of formed triangles and signed links, respectively. An automated estimation approach for the two parameters allows BSCL to take as input a signed network, learn appropriate parameters needed to model the key properties, and then output a similar network maintaining the desired properties. 
 
 Our future work will consist of further investigating both directed and weighted signed networks in future work. Furthermore, there have been recent advances for using deep learning approaches for enhancing signed networks~\cite{derr2018sgcn,Wang-etal2017,Wang-etal2018} tasks and constructing unsigned network models~\cite{Grover-etal2018,You-etal2018,Li-etal2018,de2018molgan}. Thus we will also pursue the use of different deep learning architectures to construct a deep generative model specific to signed networks.
\appendix
%Appendix A
\section{Appendix}
\subsection{Distribution of Triangle Types}
In Table~\ref{tab:triangle_type_dist} we had shown a fine-grained comparison of the distribution of signed triangle types for the Bitcoin-OTC dataset. Here, for completeness, we provide similar tables for the other two datasets. More specifically, in Table~\ref{tab:triangle_type_dist_bitcoin_alpha} we present the triangle distribution for networks generated from the Bitcoin-Alpha dataset, and Table~\ref{tab:triangle_type_dist_epinions} contains the results from the Epinions dataset.

\begin{table}[]
             \begin{center}
             \small
             \vskip -1em
             \caption{\label{tab:triangle_type_dist_bitcoin_alpha} Distribution of Triangle Types (Bitcoin-Alpha).}
             \begin{tabular}{|c|c|c|c|c|c|c|c|c|}	\hline
Triad Type			&	Real	&	STCL/SKron	&	Ants14	&	Evo07	&	BSCL	\\ \hline
\{+,+,+\}			&	0.793	&	0.766		&	0.446	&	0.750	&	0.804	\\ 
\{+,+,-\}			&	0.134	&	0.213		&	0.200	&	0.250	&	0.174	\\ 
\{+,-,-\}			&	0.069	&	0.020		&	0.341	&	0.000	&	0.021	\\ 
\{-,-,-\}			&	0.004	&	0.001		&	0.013	&	0.000	&	0.001	\\ \hline
Absolute Difference & 			&	0.158		&	0.694	&	0.232	&	0.102 \\ \hline
\end{tabular}
               \vspace{-0.05in}
               \end{center}
\end{table}

\begin{table}[t]
             \begin{center}
             \small
             \vskip -1em
             \caption{\label{tab:triangle_type_dist_epinions} Distribution of Triangle Types (Epinions).}
             \begin{tabular}{|c|c|c|c|c|c|c|c|c|}	\hline
Triad Type		&	Real	&	STCL/SKron&	Ants14	&	Evo07	&	BSCL	\\ \hline
\{+,+,+\}		&	0.808	&	0.572		&	0.929	&	0.587	&	0.698	\\ 
\{+,+,-\}		&	0.096	&	0.351		&	0.061	&	0.349	&	0.250	\\ 
\{+,-,-\}		&	0.084	&	0.072		&	0.010	&	0.052	&	0.050	\\ 
\{-,-,-\}		&	0.013	&	0.005		&	0.000	&	0.012	&	0.002	\\ \hline
Absolute Difference & 		&	0.511		&	0.243	&	0.507	&	0.309 \\ \hline
\end{tabular}
               \vspace{-0.05in}
               \end{center}
\end{table}

\begin{acks}
The authors wish to thank the anonymous reviewers for their helpful comments. Tyler Derr and Jiliang Tang are supported by the National Science Foundation
(NSF) under grant number IIS-1714741 and IIS-1715940.
\end{acks}

\bibliographystyle{ACM-Reference-Format}
\bibliography{signed_network_modeling}

\end{document}